\documentclass[rsi,superscriptaddress,twocolumn,amsmath,amssymb]{revtex4}

\usepackage{color, units, graphicx, braket}

\begin{document}
\title{Optimally focused cold atom systems obtained using density-density correlations}

\author{Andika Putra}
\affiliation{Joint Quantum Institute, University of Maryland, and National Institute of Standards and Technology, College Park, Maryland, 20742, USA}
\author{Daniel L. Campbell}
\affiliation{Joint Quantum Institute, University of Maryland, and National Institute of Standards and Technology, College Park, Maryland, 20742, USA}
\author{Ryan M. Price}
\affiliation{Joint Quantum Institute, University of Maryland, and National Institute of Standards and Technology, College Park, Maryland, 20742, USA}
\author{Subhadeep De}
\affiliation{Joint Quantum Institute, University of Maryland, and National Institute of Standards and Technology, College Park, Maryland, 20742, USA}
\affiliation{CSIR-National Physical Laboratory, New Delhi 110012, India}
\author{I. B. Spielman}
\affiliation{Joint Quantum Institute, University of Maryland, and National Institute of Standards and Technology, College Park, Maryland, 20742, USA}

\date{\today}

\begin{abstract}
Resonant absorption imaging is a common technique for detecting the
two-dimensional column density of ultracold atom systems. In many
cases, the system's thickness along the imaging direction greatly
exceeds the imaging system's depth of field, making the identification
of the optimally focused configuration difficult. Here we describe
a systematic technique for bringing Bose-Einstein condensates (BEC) and other cold-atom systems into
an optimal focus even when the ratio of the thickness to the depth of field is large: a factor of 8 in this demonstration with a BEC.  This technique relies on defocus-induced artifacts in the Fourier-transformed density-density correlation function (the power spectral density, PSD). The spatial frequency at which these artifacts first appear in the PSD is maximized on focus; the focusing process therefore both identifies and maximizes the range of spatial frequencies over which the PSD is uncontaminated by finite-thickness effects.
\end{abstract}

\maketitle

\section{Introduction}

Since the most important technique for obtaining properties of ultracold atoms is direct
imaging, a well-designed and well-aligned imaging system is crucial
for obtaining high quality data which is valid at all length scales.  While large scale properties such as the system's width or peak density can be obtained with little effort, significant care must be taken for experiments requiring very good spatial resolution \cite{Bakr2010,Sherson2010}, or those studying correlations \cite{Folling2005,Hung2011}. It is difficult to bring objects extended along the imaging axis, such as degenerate Fermi gases \cite{GranadePRL02,Regal03}, 3D Mott insulators \cite{Greiner2002}, and Bose-Einstein condensates (BECs) \cite{Anderson1995,Davis1995a}, into focus particularly after time-of-flight (TOF) expansion because their spatial thickness often exceeds the imaging system's depth of field.
Even for such objects, a high degree of accuracy in focusing is required to minimize imaging artifacts.  Understanding and minimizing these artifacts is particularly important when studying density-density correlations, where the artifacts can be confused with the correlation signal under study \cite{Hung2011,Choi2012,Seo2013,Langen2013,De2012}. Here we describe a fairly generic technique for focusing on these extended objects which is far more precise than simply optimizing the ``sharpness'' of imaged atom clouds.


Absorption imaging is a ubiquitous approach for measuring the density distribution of ultracold atom systems ~\cite{Ketterle1999}. A probe beam illuminates the atomic system, and the resulting shadow is imaged onto a scientific camera, typically a charge-coupled device (CCD) or complementary metal-oxide-semiconductor (CMOS) detector. Ideally, the fraction of light absorbed would be directly related to the two-dimensional column density $\rho_{\rm 2D} (x,y)=\int {\rm d}z\;\rho(x,y,z)$ of the atoms along the imaging direction ${\bf e}_z$, where $\rho(x,y,z)$ is the density of atoms. If the thickness $\delta z$ along ${\bf e}_z$ exceeds the imaging system's depth of field, then some of the atomic distribution must necessarily be out of focus, invalidating any simple relationship between absorption and column density.  Given this, it is a challenge to obtain the optimal focal plane of the extended system that minimizes the artifacts resulting from this defocus, e.g., at the center of a distribution symmetric along ${\bf e}_z$.

Typically a system is brought into a focus by minimizing the size or apparent diffraction effects from a compact object such as a trapped BEC; in many cases, no such compact reference at the desired image plane is available.  In this paper, we present a technique for determining the optimal focus of absorption-imaged extended objects. Using this technique, we identify the focal plane within an accuracy of $\unit[2]{\mu m}$ for a $\delta z = \unit[150]{\mu m}$ thick object. Specifically, given an object with density-density correlations~\cite{Hung2011} with a spatial correlation length $\ell$, we show that observations of correlations in the optical absorption as a function of camera position allow us to bring the object into focus to within a fraction of the depth of field associated with $\ell$, even without knowing the details of the correlation function. This optimal focus is the camera position where the imaged auto-correlation function (ACF) most accurately reflects the atomic density-density correlations, minimizing both defocus-induced artifacts and the resolution limiting effect of the system's finite thickness \cite{Langen2013,De2012}.

In this paper, we review the basic theoretical formulation required to
understand light propagating through an absorbing dielectric medium.  We then consider
several example images created by different idealized objects, in each case noting how to determine their optimal focus. Lastly, we experimentally apply this technique to images of BECs after TOF.

\section{Theory}

Monochromatic light of free-space wavelength $\lambda$ and wavenumber $k_0 = 2\pi/\lambda$ propagating through an object with complex relative permittivity $\varepsilon(\mathbf{r})=\epsilon/\epsilon_0$ and relative susceptibility $\chi({\bf r}) = \varepsilon(\mathbf{r})-1$, where $\epsilon$ is the permittivity, and $\epsilon_0$ is the electric constant, is described by the vectorial wave equation for the electric field $\mathbf{E(r)}$:
\begin{equation}
\label{Eq1:vectorial}
\nabla^2 \mathbf{E(r)}+{k_0^2}\varepsilon(\mathbf{r}) \mathbf{E(r)}=-\nabla\left[ \mathbf{E(r)}\cdot \nabla \ln \varepsilon(\mathbf{r})\right].
\end{equation}
In a medium where $\varepsilon(\mathbf{r})$ is slowly varying, the right-hand side (rhs) of Eq. \eqref{Eq1:vectorial} can be neglected, reducing Eq. \eqref{Eq1:vectorial} to separate scalar wave equations
\begin{equation*}
\label{Eq2:scalar}
\nabla^2 E\mathbf{(r)}+k_0^2 \varepsilon(\mathbf{r}) E\mathbf{(r)}=0,
\end{equation*}
for each vector component of $\mathbf{E(r)}$, e.g., we might have $\mathbf{E(r)}=E\mathbf{(r)} \mathbf{e}_x$ for light linearly polarized along $\mathbf{e}_x$.

\subsection{Wavefield propagation}

Here, we cast the above scalar wave equation into the form
\begin{eqnarray}
\label{Eq2:scalar_z}
-\frac{\partial^2 E\mathbf{(r)}}{\partial z^2} &=& \left [\nabla^2_\perp + k_0^2 \right] E\mathbf{(r)} +k_0^2 \chi({\bf r}) E(\mathbf{r}),
\end{eqnarray}
suitable for light predominantly traveling along ${\bf e}_z$. For a known field configuration at $E({\bf r})$ (such as the probe laser before it interacts with the atoms), Eq. \eqref{Eq2:scalar_z} has the formal solution
\begin{equation}
\label{Eq3:general_sol}
E\left({\bf r}+\Delta z {\bf e}_z\right)=\exp\left[\pm i\Delta z\sqrt{\nabla^2_\perp+k_0^2 + k_0^2 \chi({\bf r})}\right]\!E\left({\bf r}\right),
\end{equation}
describing the field propagated a distance $\Delta z$ along ${\bf e}_z$. [plus minus sign]

Wave propagation in free space [i.e., $\chi({\bf r})=0$ in Eq.\eqref{Eq2:scalar_z}] is solved exactly in the angular spectrum representation ~\cite{novotny06}
\begin{eqnarray}
\label{angspec}
E_{\rm fs}({\bf r}\!+\!\Delta z {\bf e}_z) &\!\! =\!\! &\mathbf{P}(\Delta z)E({\bf r})\\ &\!\! =\!\!\! & \int\!\mathrm{d}^2\mathbf{k}_{\rm 2D}\! \left[\tilde{{\bf P}}(\mathbf{k}_{\rm 2D},\Delta z) \tilde{E}(\mathbf{k}_{\rm 2D},z) \right]\! e^{i {\bf k}_{\rm 2D}\cdot\mathbf{r}_{\rm 2D}},\nonumber
\end{eqnarray}
for a forward going wave, 
with the 2D position ${\bf r}_{\rm 2D} = \left(x,y\right)$ and wavevector $\mathbf{k}_{\rm 2D}=\left(k_{x},k_{y}\right)$; the Fourier-transformed wavefield  $\tilde{E}({\bf k}_{\rm 2D},z)=\int {\rm d}^2 {\bf r}_{\rm 2D} \:\exp(-i  {\bf k}_{\rm 2D}\cdot\mathbf{r}_{\rm 2D}) E(\mathbf{r})$; and the transfer function for propagating a distance $\Delta z$ in free space
\begin{equation*}
\tilde{{\bf P}}\left({\bf k}_{\rm 2D},\Delta z\right)=\exp\left[i\Delta z\left(k_0^{2}-k^2_{\rm 2D}\right)^{1/2}\right].
\end{equation*}
The transfer function behaves differently in two regions of spatial
frequencies: for $k^2_{\rm 2D}<k_0^{2}$, $\tilde{\bf P}$ is oscillatory (propagating regime), and for $k^2_{\rm 2D}>k_0^{2}$, it is exponentially decaying (evanescent regime).

Meanwhile, considering only $\chi ({\bf r})$ [neglecting the first term in the rhs of Eq. \eqref{Eq2:scalar_z}], the absorption and refraction of light traveling a distance $\Delta z$ is described by
\begin{eqnarray}
\label{BLlaw}
E_{\rm BL}({\bf r}\!+\!\Delta z {\bf e}_z) & = &\mathbf{Q}(\Delta z) E({\bf r}) \\
 & = & \exp \left[ i k_0 \int_z^{z+\Delta z}\mathrm{d}z \sqrt{\chi({\bf r})}\right] E({\bf r}).\nonumber
\end{eqnarray}
Unlike the usual Beer-Lambert (BL, discussed in Sect.~\ref{sect:BL}), this expression alone does not reflect a good approximation to beam propagation for systems of any significant thickness.

\subsection{Beer-Lambert law and the paraxial approximation}\label{sect:BL}

To better understand the independent influence of the beam's propagation and its interaction with matter, we apply the paraxial approximation to Eq.~\eqref{Eq2:scalar_z}, allowing us to draw an analogue between the paraxial wave equation and the Schr\"odinger equation, which can be solved numerically using a split-step Fourier method (SSFM)~\cite{Fleck95}. To understand the difference between Eq.~\eqref{BLlaw} and the usual BL law, we again turn to Eq.~\eqref{Eq2:scalar_z}, now assuming that the electric field can be written as $E({\bf r}) = \exp\left(i k_0 z\right) E^\prime({\bf r})$, where $E^\prime({\bf r})$ is a slowly varying envelope along ${\bf e}_z$.  Inserting this form into Eq.~\eqref{Eq2:scalar_z} gives the paraxial wave equation
\begin{eqnarray}
\label{Eq5:parax}
-2 i k_0\frac{\partial E'\mathbf{(r)}}{\partial z} &=& \nabla^2_\perp E'\mathbf{(r)} +k_0^2 \chi({\bf r}) E'(\mathbf{r}),
\end{eqnarray}
where the assumed weak $z$ dependence of $E^\prime({\bf r})$ allowed us to drop $\partial^2_z E^\prime({\bf r})$.  Like above, the spatial evolution of an initial $E^\prime({\bf r})$ can be partitioned into a spectral part $\tilde{\bf P}^\prime({\bf k}_{\rm 2D},\Delta z)$ and a coordinate part  ${\bf Q}^\prime(\Delta z)$, with
\begin{align}
\label{eq:paraxial1}
\tilde{\bf P}^\prime({\bf k}_{\rm 2D},\Delta z) &= \exp\left(-i\frac{k_{\rm 2D}^2}{2 k_0} \Delta z \right)\\
\label{eq:paraxial2}
{\bf Q}^\prime(\Delta z) &= \exp \left[ i \frac{k_0}{2} \int_z^{z+\Delta z} \chi({\bf r})\mathrm{d}z\right].
\end{align}
For the paraxial approximation to be valid, the condition $\left|\chi({\bf r})\right|\ll1$ must also hold: otherwise the ${\bf Q}^\prime(\Delta z)$ evolution would lead $E^\prime({\bf r})$ to depend strongly on $z$.

We numerically evolve the paraxial wave equation [Eq. \eqref{Eq5:parax}] along ${\bf e}_z$ using a split-step Fourier method (SSFM)~\cite{Korpel1986,Feit78}, where the operators in the rhs of Eq. \eqref{Eq5:parax} are split into two: one operator represents wave propagation in a uniform medium using Eq. \eqref{eq:paraxial1} and the other operator takes into account the effect of refractive index variation using Eq. \eqref{eq:paraxial2}.  In the SSFM, we alternately apply the two evolution operators with steps of size $\Delta z$. For each step, the complex amplitude $E'(\mathbf{r})$ is propagated first by $\mathbf{P}'(\Delta z /2)$, then by $\mathbf{Q}'(\Delta z)$, and then again by $\mathbf{P}'(\Delta z/2)$. The resulting symmetrized split evolution
\begin{equation*}
\label{Eq3:SSFM}
E'\left({\bf r}+\Delta z {\bf e}_z\right)=\mathbf{P'}\left(\Delta z/2\right)\mathbf{Q'}(\Delta z)\mathbf{P'}\left(\Delta z/2\right)E'\left({\bf r}\right),
\end{equation*}
has its first correction at order $\Delta z^3$.

The paraxial equations allow us to introduce the {\it depth of field}
\begin{equation}
\label{eq:dof}
d_{\rm dof} = \frac{2 k_0}{k_{\rm max}^2} = \frac{l_{\rm min}^2}{\pi \lambda},
\end{equation}
where $k_{\rm max}^2$ is the largest $k_{\rm 2D}$ of interest and $l_{\rm min} = 2\pi/k_{\rm max}$ is the corresponding minimum length scale [these might be specified by: the maximum significant wavevector in $\chi({\bf k}_{\rm 2D},z)$; the resolution of the physical imaging system; or at most by $k_0$].

We obtain the BL law by assuming that the system is thin along ${\bf e}_z$, i.e., both $\delta z\ll d_{\rm dof}$, {and $\tilde{\bf P}^\prime({\bf k}_{\rm 2D},\Delta z)$ may be neglected.  For purely absorbing materials where $\chi({\bf r}) \propto i \sigma_0\rho({\bf r})$, this gives the usual BL law
\begin{eqnarray}
\label{BLLawIntensity}
I({\bf r}+\delta z {\bf e}_z) & = & \exp\left[ - \sigma_0 \int_z^{z+\delta z} \rho({\bf r})\mathrm{d}z\right]I({\bf r})
\end{eqnarray}
describing the attenuation of the free space optical intensity $I({\bf r}) = c \epsilon_0\left|E({\bf r})\right|^2/2$ by absorbers of density  $\rho({\bf r})$ and scattering cross-section $\sigma_0$.  This BL result can also be obtained without the paraxial approximation by first neglecting the $\nabla^2_\perp$ term in Eq.~\eqref{Eq2:scalar_z} (valid when $k_{\rm max} \delta z \ll 1$: a more strict requirement than in the paraxial approximation where we had $\delta z \ll d_{\rm dof}$) and again assuming $\left|\chi({\bf r})\right|\ll1$, a small relative susceptibility~\footnote{Moreover, the gradient term $\nabla \ln \varepsilon(\mathbf{r})$ in Eq. \eqref{Eq1:vectorial} cannot be safely neglected for systems where $\left|\chi({\bf r})\right|$ is large or sufficiently rapidly varying, although this would generally imply a breakdown of the paraxial approximation as well.}.

In experiment, the BL law is generally applied by comparing the intensities $I({\bf r}_{\rm 2D})$ and $I_0({\bf r}_{\rm 2D})$ measured with and without atoms present, respectively.  This relates the optical depth
\begin{equation*}
{\rm OD}({\bf r}_{\rm 2D}) \equiv -\ln\frac{I({\bf r}_{\rm 2D})}{I_0({\bf r}_{\rm 2D})} = \sigma_0 \rho_{\rm 2D}({\bf r}_{2D})
\end{equation*}
to the 2D column density.  In cold atom experiments, this column density is the primary observable in experiment.

\subsection{Absorption Imaging}

\begin{figure}
\includegraphics[width=3.3in]{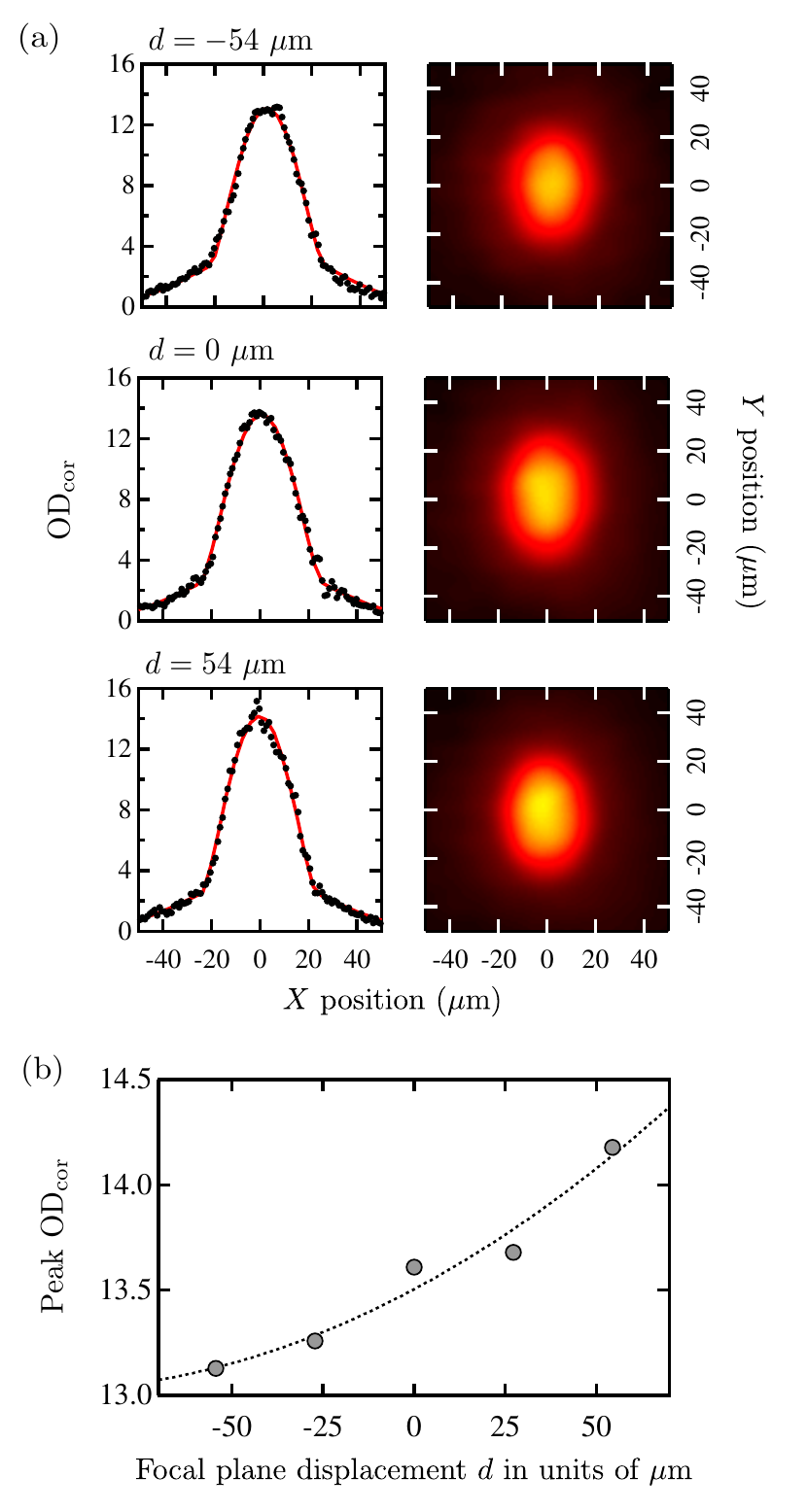}
\caption{Dependence of BEC images on image plane position.  (a) Intensity corrected optical depth measured $d=-\unit[54]\mu{\rm m}$, 0 $\mu{\rm m}$, and $\unit[54]\mu{\rm m}$ from the optimal focus: images (right) and line cuts at $y=0$ (left).  (b) The peak optical depth depends only weakly on $d$; is not maximized at $d=0$; and has no structure on the $\unit[2]{\mu m}$ scale.}
\label{Fig:BECs}
\end{figure}

Here we consider systems of ultracold atoms illuminated by laser light on a cycling transition, where the atom-light interaction is described by a complex relative susceptibility 
\begin{equation*}
\chi({\bf r})  =  \frac{\sigma_{0}}{k_0}\left[ \frac{i-2\delta/\Gamma}{1+{I}/{I_{\rm sat}} +\left( {2\delta}/{\Gamma} \right)^2}\right]\rho(\mathbf{r}).
\end{equation*}
$\rho(\mathbf{r})$ is the atomic density; $\delta$ is the laser's detuning from atomic resonance;  $\sigma_{0} = 6\pi/k_0^2$ is the resonant scattering cross-section; $\Gamma$ is the atomic linewidth; and $I_{\rm sat}$ is the saturation intensity~\cite{Ketterle1999}.  The standard BL law is valid for dilute ($\rho\ll k_0^3$, see Ref.~\footnote{A dilute regime, with density $\rho \ll k_0^3$, can be achieved by letting the atomic cloud to ballistically expand.}), spatially thin systems ($k_0 \delta z \ll 1$), illuminated by low intensity  ($I_0 \ll I_{\rm sat}$) probe beams.

The $I_0 \ll I_{\rm sat}$ requirement can be lifted by introducing the intensity-corrected optical depth
\begin{equation}
\label{Eq:OD}
{\rm {OD_{cor}}}({\bf r}_{2D}) \equiv - \ln\frac{I({\bf r}_{2D})}{I_{0}({\bf r}_{2D})}+\frac{I_0({\bf r}_{2D})-I({\bf r}_{2D})}{I_{\rm sat}},
\end{equation}
which is related to the column density 
\begin{equation}
\label{Eq:rho2D}
\rho_{\rm 2D}({\bf r}_{2D})=\frac{{\rm {OD_{cor}}}({\bf r}_{2D})}{\sigma_0},
\end{equation}
of dilute ($\rho\ll k_0^3$), spatially thin systems ($k_0 \delta z \ll 1$).  Due to the limited dynamic range of the camera's pixels \cite{reinaudi07} and the presence of background light, it is technically difficult to reliably detect uncorrected optical depths, larger than $\approx 4$. Thus, we deliberately select $I_0>I_{\rm sat}$, saturating the transition with $I_0$ such that ${\rm OD_{cor}}<3$.

In addition, the spatial thickness of many cold atom systems exceed the depth of field leaving parts of its distribution along imaging direction inevitably out of focus, thereby invalidating Eq. \eqref{Eq:rho2D}.  Even for dilute clouds (after sufficient TOF), images taken an equal distance above and below the focal plane can differ.  This lack of symmetry makes a straightforward determination of the optimal focus difficult (lensing effects from even slightly off-resonance imaging beams and aberrations in the imaging system can complicate the situation further.)

To illustrate this difficulty, we consider images of BECs with the focal plane displaced a distance $d=\unit[-54]{\mu m}$, 0 ${\rm \mu m}$, and $\unit[54]{\mu m}$ from the BECs' center (see Fig.~\ref{Fig:BECs}).  Because the BEC is thick compared to the depth of field, Eq. \eqref{Eq:rho2D} does not hold; in addition lensing effects cause the cloud's peak $\rm{OD_{cor}}$ to behave asymmetrically when the focus is behind or in front of the cloud.  In these images, there are no sharp features that identify the optimal focus at the micron level. Owing to the weak dependence of large-scale parameters such as peak-height or width on defocus, such precise focusing is not required in many experiments.  As we see below, experiments that study correlations within such images are extremely sensitive to defocus and new methods are required.  Our technique brings images such as these into focus, identifying an optimal focal plane at the $\approx \unit[2]{\mu m}$ level.

\subsection{Modeling}

To obtain a basic understanding of our approach, we first consider the defocused image of a $\unit[1]{\mu m}$ thick absorbing medium, inhomogenous ${\bf e}_x$-${\bf e}_y$ plane, bounded above and below by vacuum, with, $\chi({\bf r})=i g(x,y)$ for $z\in(\unit[-0.5]{\mu m}, \unit[0.5]{\mu m})$, where $g(x,y) \geq 0$ is a Poisson distributed random variable.  Like atoms illuminated on resonance, this medium has a purely imaginary susceptibility.  The illuminating light is modeled by a plane wave with wavelength $\lambda=\unit[780]{nm}$ suitable for imaging our $^{87}$Rb Bose-Einstein condensates~\footnote{In our SSFM simulation of light traversing this medium, we used a $\Delta z = \unit[1]{\mu m}$ step size.}.  While this object has no visible structure, by virtue of its spectrally flat density-density correlation function, it can be brought into focus.

The imaged intensity pattern $I(x,y)$ from this $\unit[1]{\mu m}$ layer appears random at various distances from focus, but its correlations become oscillatory.  To reveal this information, we turn to its spatial power spectral density: the magnitude squared of $I(x,y)$'s Fourier transform \footnote{The Wiener-Khincin theorem states that the spectral decomposition of the autocorrelation function is equal to the power spectral density.}.  The PSD  is circularly symmetric in the spatial frequency $\mathbf{k}_{\rm 2D}=(k_x,k_y)$ plane. Fig.~\ref{Fig:theory}a shows the PSD in this $k=\left|\mathbf{k}_{\rm 2D}\right|$ ``radial'' direction as a function of distance from focus $d$. This PSD has a fringe pattern; the wavevector of the first minimum exceeds the maximum imaged wavevector only near the image's focus at $d=0$ ${\mu m}$.

The physical origin of this structure can be understood by turning to the paraxial wave equations [Eqs.~\eqref{eq:paraxial1} and \eqref{eq:paraxial2}], and by first studying a single absorber at ${\bf r}=0$ illuminated by a plane wave $E_0^\prime({\bf r}_{\rm 2D},0^-) = E_0$.   Equation~\eqref{eq:paraxial2} shows that a thin absorber simply changes the amplitude of the field, leaving its phase untouched, and for simplicity, we assume this absorber has a Gaussian profile in the ${\bf e}_x$-${\bf e}_y$ plane with width $w_0$.  Thus the electric field just following the absorber is changed by $\delta E^\prime({\bf r}_{\rm 2D},0^+) = -\delta E \exp\left[ -r^2/w_0^2\right]$, with $r^2 = x^2 + y^2$.  The propagation of such a gaussian mode by a distance $d$ along ${\bf e}_z$ can be solved exactly in the paraxial approximation, and in the spectral basis this is
\begin{align*}
\delta \tilde E^\prime({\bf k}_{\rm 2D},d) &= - \pi w^2_0\delta E \exp\left[ -\frac{w_0^2 k_{\rm 2D}^2}{4}\left(1+\frac{2i}{w_0^2 k_0} d\right) \right].
\end{align*}
The total field from an absorber located at a different location ${\bf r}_0$ in the ${\bf e}_x$-${\bf e}_y$ plane simply acquires an overall phase factor $\exp\left[-i {\bf k}_{\rm 2D}\cdot{\bf r}_0\right]$. We now compute the experimentally relevant optical depth by taking the reverse Fourier transform of the full electric field, computing the intensity, then the optical depth, and taking the Fourier transform to obtain (retaining terms of order $\delta E / E_0$) 
\begin{align}
\label{eq:singleabs}
\widetilde{\rm OD} &= \frac{2 \pi w_0^2 \delta E}{E_0}\exp\left( -\frac{w_0^2 k_{\rm 2D}^2}{4}\right)\cos\left(\frac{k_{\rm 2D}^2 d}{2 k_0}\right),
\end{align}
with the same overall phase factor depending on the initial position.  Averaging over $N$ randomly placed absorbers therefore gives an overall signal scaling as $\sqrt{N}$ with a random overall phase.  Taking the magnitude squared gives the PSD
\begin{align*}
{\rm PSD}_{\rm thin} &= {\rm PSD}_0\times\cos^2\left(\frac{k_{\rm 2D}^2 d}{2 k_0}\right),
\end{align*}
with
\begin{align*}
{\rm PSD}_0 &= N\left(\frac{2 \pi w_0^2 \delta E}{E_0}\right)^2\exp\left( -\frac{w_0^2 k_{\rm 2D}^2}{2}\right).
\end{align*}

This quantity has zeros located at $k_{\rm zero}[n] = \sqrt{2\pi(n+1/2) k_0 / d}$ for integer $n$.    In our numerical simulation, the minima follow the functional form $k_{\rm zero}[n] = A[n]|d|^{-1/2}$ as shown by the dotted lines in Fig.~\ref{Fig:theory}, with $A[0]\approx5.08$ and $A[1]\approx8.76$ for the first and second zeros: the expected values for $A[n]$.  Thus, for this thin apparently structureless system, fringes in the PSD allow us to identify the focal plane.

\begin{figure}
\includegraphics[width=2.5in]{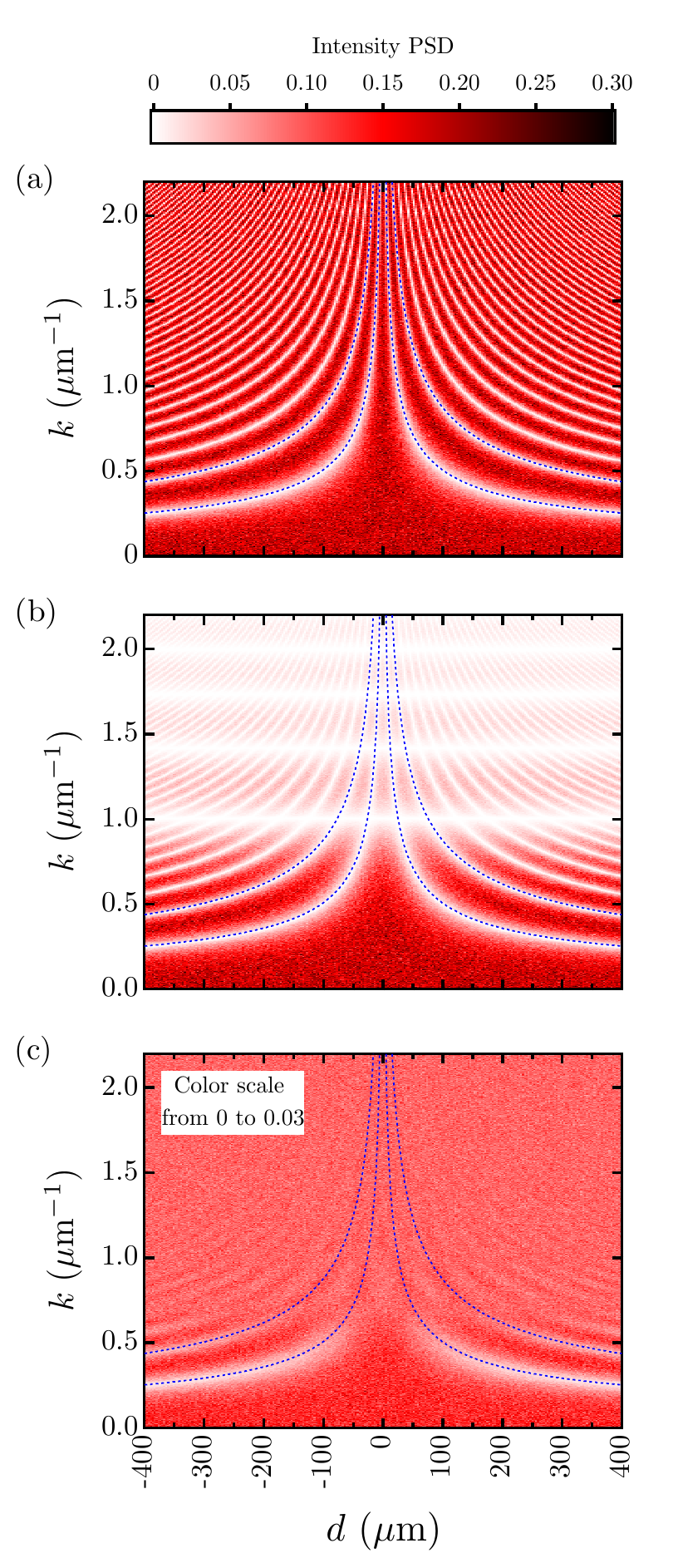}
\caption{(a) Spatial PSD of the intensity produced by $\unit[1]{\mu m}$ thick layer of randomly distributed scatterers showing that fringes diverge in focus.
(b) PSD produced by a $\unit[100]{\mu m}$
thick sheet of random columnar scatterers. (c) PSD produced by a $\unit[100]{\mu m}$ thick sheet of random scatterers.  The dotted lines are functional forms of the lowest curved-fringes, and in each case $d$ is measured from the objects' center.}
\label{Fig:theory}
\end{figure}

To demonstrate the technique of finding optimal focus of an extended object,
we now consider a second disordered scattering potential with a columnar structure, now $\unit[100] {\mu m}$ thick, i.e., $\chi({\bf r})=i g(x,y)$ for $z\in(\unit[-50]{\mu m}, \unit[50]{\mu m})$, where again $g(x,y) \geq 0$ is a Poisson distributed random variable. This object's PSD is plotted as a function of distance $d$ from its center in Fig.~\ref{Fig:theory}b; in addition to the same fringe pattern as for the $\unit[1]{\mu m}$ thick case, the PSD now vanishes at specific spatial frequencies independent of $d$.  To model this, we note that the absorbers can now be located at a distance $z$ from the symmetry plane, so in Eq.~\eqref{eq:singleabs}, we replace $d\rightarrow d-z$ and integrate $z$ from $-\delta z/2$ to $\delta z/2$, which ultimately gives the PSD
\begin{align*}
{\rm PSD}_{\rm col} &= {\rm PSD}_0 \times \cos^2\left(\frac{k_{\rm 2D}^2 d}{2 k_0}\right) {\rm sinc}^2\left(\frac{k_{\rm 2D}^2 \delta z}{4 k_0}\right).
\end{align*}
This predicts the appearance of additional zeros located at $k^\prime_{\rm zero}[m] = \sqrt{4\pi m k_0/\delta z}$ for non-zero integer $m$ (this is an artifact of the box-like density distribution of atoms, and would be greatly softened in real systems where the density drops smoothly to zero).  In our example, the lowest order horizontal fringes is located at $k^\prime_{\rm zero}[1]=\unit[1.00]{\mu m^{-1}}$. Here again, we easily determine the optimal focus, $d=0\ {\mu m}$, from the diverging curved-fringes. 

Next, we consider a scattering potential fully disordered in 3D, again with a $\unit[100]{\mu m}$ thickness,  i.e., $\chi({\bf r})=i g(x,y,z)$ for $z\in(\unit[-50]{\mu m}, \unit[50]{\mu m})$, where $g(x,y,z) \geq 0$ is a Poisson distributed random variable.  In this case, the independent random scatterers along imaging direction causes the PSD to
rapidly loose structure with increasing $k_{\rm 2D}$ (see Fig.~\ref{Fig:theory}c).  Here too, our random scatter model can be applied, giving 
\begin{align*}
{\rm PSD}_{\rm rnd} &= {\rm PSD}_0 \times \left[\frac{\cos\left(k_{\rm 2D}^2 d/k_0\right) {\rm sinc}\left(k_{\rm 2D}^2 \delta z/2 k_0\right)+1}{2}\right].
\end{align*}
This reduces to our earlier result when $\delta z\rightarrow 0$ for a thin system and shows that, while the same fringes exist, they are rapidly attenuated for larger spatial frequencies, where the signal approaches a constant background value.  However, in principle the curved-fringes still allow the optimal focus to be identified. 

\section{Optimal focusing of Elongated Bose-Einstein Condensates}

Using on our model, we now consider absorption
imaged BECs and implement the technique presented in previous section
to find the optimal focus.

\begin{figure}
\centering{}
\includegraphics[width=2.7in]{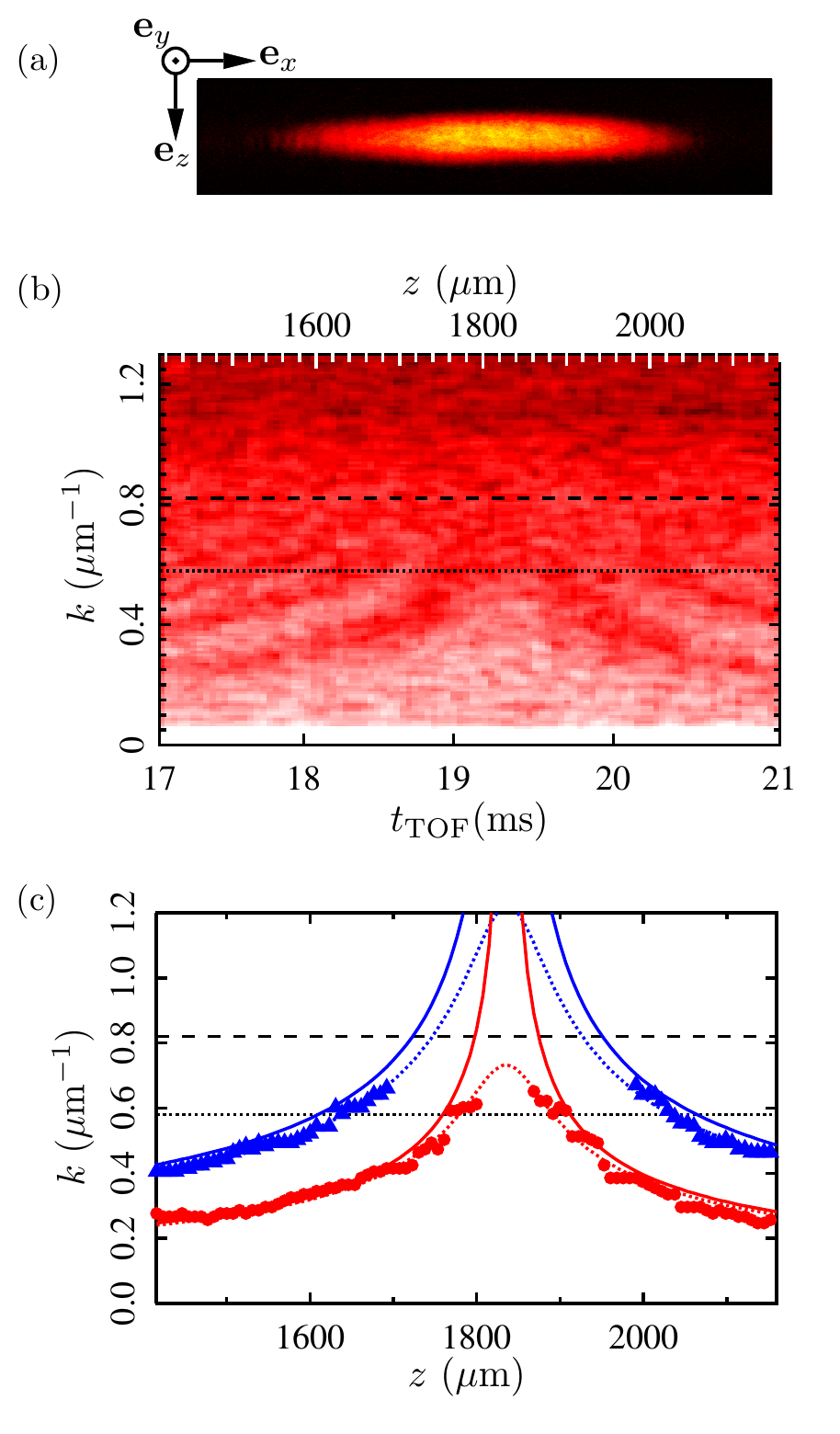}
\caption{(a) Absorption imaged elongated BEC with density fluctuations. (b) 1D PSD of column density along weakly trap direction ${\bf e}_x$ as a function of $t_{\rm TOF}$. (c) Values of $k$ where the 1D PSD is minimum. The two lowest such $k$-fringes are depicted. Symbols denote the fringe locations extracted from (b) plotted along with Lorentzian fits (dotted lines), determining the optimal focus. The solid curves depict theoretical functional forms for the two lowest order fringes. In (b) and (c) the dashed line marks $k = k^\prime_{\rm zero}[1]=\unit[0.82]{\mu m^{-1}}$ for our condensate thickness of $\unit[150]{\mu m}$; the dotted line marks $k = k^\prime_{\rm zero}[1]/\sqrt{2}$, below which the ACF of the focused images reliably reflects the ACF $n_{\rm 2D}({\bf r}_{\rm 2D})$.}
\label{fig:data}
\end{figure}

We prepared $N=7\times10^5$ atom $^{87}$Rb Bose-Einstein condensates in the $\left|5\mbox{S}_{1/2},F=1, m_F=0\right\rangle$ electronic ground state in a crossed-dipole trap with frequencies $\omega_{x,y,z}=2\pi\times\unit[(3.1,135,135)]{Hz}$. In situ, the BECs were javelin shaped owing to the extremely anisotropic confining potential. After a 17 ms to 21 ms TOF, we repumped into the $f=2$ manifold, and resonantly imaged on the $\left|5\mbox{S}_{1/2},f=2,m_F=2\right\rangle$ to $\left|5\mbox{P}_{3/2},f=3,m_F=3\right\rangle$ cycling transition with a $\lambda\approx\unit[780.2]{nm}$ probe laser.

The imaging system consisted of a CCD camera and two pairs of lenses functioning as a compound microscope, magnifying the intensity pattern at the object by a factor of $\approx6$  at the image plane. The first pair of objective lenses, with effective focal length (efl) $\unit[\hat{f_1}=53.6]{mm}$, collimated the light diffracted by the cloud and were separated by a distance $D=\hat{f_1}+\hat{f_2}$ from a second pair of lenses with a $\hat{f_2}=\unit[325]{mm}$ efl.  The resulting $0.23$ numerical aperture implies that a $\unit[10.6]{\mu m}$ diffraction-limited spot on our CCD sensor is larger than its $\unit[5.6]{\mu m}$ pixel size. The associated $\unit[1.7]{\mu m}$ spot-size on the cloud gives a $d_{\rm dof} = \unit[18.6]{\mu m}$ depth of field in our imaging system \cite{inoue1995}.

Instead of varying the distance from focus by physically moving imaging lenses or the CCD, we changed the time during which the BEC fell along ${\bf e}_z$ and obtained absorption images with TOF times $t_{\rm TOF}$ from $\unit[17.0]{ms}$ to $\unit[21.0]{ms}$.  At these TOFs, the condensates' radii were $R_{y,z}\approx\unit[75(5)]{\mu m}$
and $R_{x}\approx\unit[210(10)]{\mu m}$. Initially, the cloud was elongated in the harmonic trap with aspect ratio $43$ to 1.  The initial $43:1$ aspect ratio was reduced to $2.65:1$ after TOF, and the transverse size of the cloud exceeded the imaging depth of field by a factor of $8$.

Figure~\ref{fig:data}b shows the 1D PSD of the atoms' corrected optical depths along ${\bf e}_z$, which is directly related to the absorption intensity through Eq. \eqref{Eq:OD}.  The fluctuations in the BEC's density distribution behave like the randomly modulated $\chi({\bf r})$ in our example systems, creating a recurring fringe pattern in the PSD spectrum
as obtained in Fig.~\ref{fig:data}c. The fringes are quite pronounced for quasi one-dimensional BECs, where initial phase fluctuations map into pancake-shaped density fluctuations arrayed along the initially long axis after TOF~\cite{dettmer01}. Despite the decreased contrast at high spatial frequencies due to the BEC extent along ${\bf e}_z$, we clearly observe fringes curving as a function of $t_{\rm TOF}$ in Fig. 3c. This allows us to determine the optimal focus of the system.

From the above experimental data, we fit the two lowest order fringes to $k_m \left[(d-z_0)^2/\delta z^2 + 1\right]^{-1/4}$, a peaked function with the expected $d^{-1/2}$ behavior away from $z_0$. The fits give an optimal focus location of $z_0[0]=\unit[1836(2)]{\mu m}$ using the zeroth order fringe or of $z_0[1]=\unit[1837(2)]{\mu m}$ using the first order fringe. These values correspond to a TOF of $\unit[19.36(1)]{ms}$. We are thus able to determine the optimal focus within $\approx \unit[2] {\mu m}$ or equivalently $\approx \unit[10] {\mu s}$ in TOF. Comparing the experimental data to the theoretical forms, we notice that the fringes are slightly asymmetrical with their locations slightly below theoretical ones for larger TOF. Based on our simulations, this likely results from the $z$ dependent magnification of our imaging system, which changes by about 10\% as the atoms fall from 1420 $\mu {\rm m}$ to 2150 $\mu {\rm m}$ (17 ms to 21 ms TOF).

\section{Summary}

We presented a systematic method to bring clouds of ultracold atoms,
particularly initially elongated BECs, into an optimal focus. The density fluctuations in the BECs after TOF acted like random scatterers, creating diffraction pattern which changed predictably as a function of distance from the optimal focus. Using TOF absorption imaging, we demonstrated this method, pinpointing the optimal focus of the BEC to within 2 $\mu$m for a $\unit[150] {\mu m}$ thick BEC. This robust technique is easily implemented, requires no hardware changes, and uses a minimum of computation.

\acknowledgments{We thank F.~E.~Becerra, A.~Hu, and W.~D.~Phillips for a careful reading of the manuscript.  We acknowledge the financial support from the NSF through the Physics Frontier Center at JQI, and the ARO with funds from both the Atomtronics MURI and DARPA's OLE Program.}


\end{document}